# SCALINGS FOR TOKAMAK ENERGY CONFINEMENT

P.N. YUSHMANOV[1], T. TAKIZUKA[2], K.S. RIEDEL[3],
O.J.W.F. KARDAUN[4], J.G. CORDEY[5], S.M. KAYE[6], D.E. POST[6]
International Thermonuclear Experimental Reactor*,
Max-Planck-Institut für Plasmaphysik,
Garching bei München,
Federal Republic of Germany

ABSTRACT. On the basis of an analysis of the ITER L-mode energy confinement database, two new scaling expressions for tokamak L-mode energy confinement are proposed, namely a power law scaling and an offset-linear scaling. The analysis indicates that the present multiplicity of scaling expressions for the energy confinement time $\tau_E$ in tokamaks (Goldston, Kaye, Odajima-Shimomura, Rebut-Lallia, etc.) is due both to the lack of variation of a key parameter combination in the database, $f_S = 0.32 \, R \, a^{-0.75} \, k^{0.5} \sim A \, a^{0.25} k^{0.5}$, and to variations in the dependence of $\tau_E$ on the physical parameters among the different tokamaks in the database. By combining multiples of $f_S$ and another factor, $f_q = 1.56 \, a^2 \, kB/RI_p = q_{eng}/3.2$, which partially reflects the tokamak to tokamak variation of the dependence of $\tau_E$ on q and therefore implicitly the dependence of $\tau_E$ on $I_p$ and $n_e$, the two proposed confinement scaling expressions can be transformed to forms very close to most of the common scaling expressions. To reduce the multiplicity of the scalings for energy confinement, the database must be improved by adding new data with significant variations in $f_S$, and the physical reasons for the tokamak to tokamak variation of some of the dependences of the energy confinement time on tokamak parameters must be clarified.

## 1. INTRODUCTION

Characterization of energy confinement in tokamaks is essential for developing and testing candidate theories and models for energy confinement and for identifying the parameters that should be emphasized in the design of the next generation of experiments. In the past, the difficulties in connection with characterizations based on statistical regression analyses were that, in general, each analysis usually resulted in a new scaling which fitted the existing database, but which yielded different predictions for tokamaks that were not in the database. To improve the situation, a global L-mode energy confinement time database was collected by Kaye et al. [1] and analysed in detail in Refs [1-6]. As part of the ITER joint activity, the original L-mode database of 1983 [7], containing data from DIII, ISX-B, DITE, ASDEX, TFR, PDX and PLT (also including some initial TFTR data), was updated and enlarged to include new data from the large tokamaks JT-60, TFTR, JET and DIII-D, and additional data from JFT-2M and T-10 [1]. The JT-60 and JFT-2M contributions included data from both divertor and limiter discharges.

It is difficult to characterize these data. First, tokamak transport losses are undoubtedly due to a variety of plasma, atomic and surface processes. Most scalings are expressed in terms of the parameters for heating power P, toroidal field B, plasma current $I_p$, elongation k, major radius R, minor radius a, plasma density n, and isotopic mass M. The confinement losses clearly depend on other variables as well, such as the plasma profile, current profile, heating profile, MHD behaviour and plasma–wall interactions, so that the set of eight variables normally used is incomplete. A second problem is that the functional form of the

TABLE I. VALUES OF $R^2$ FOR CANDIDATE SCALINGS [1] WITH $q < 6$ AND $P_{OH} \ll P_{aux}$

| KB | ITER89-P | G | CY | RL | ITER89-OL | OS |
|---|---|---|---|---|---|---|
| 0.98 | 0.98 | 0.975 | 0.98 | 0.94 | 0.96 | 0.92 |

KB is Kaye-Big scaling [1], ITER89-P is the ITER power law scaling discussed in this paper, G is the Goldston scaling [9], CY is the Chudnovskij-Yushmanov scaling [4], RL is the Rebut-Lallia scaling [10], ITER89-OL is the ITER offset-linear scaling described in this paper, and OS is the Odajima-Shimomura scaling [11].

dependence of the energy confinement time $\tau_E$ on the parameters is unknown. Therefore, any assumed form is probably valid only over a limited range. An example of this is the existence of both a power law scaling and an offset-linear scaling for the present tokamak database [1]. For these reasons and because of the incompleteness of the parameter set, the variation in the difference $\tau_E - \tau_E^{fit}$ is not purely due to statistical variations in the data. Moreover, there are collinearities in the data which are due to the lack of independent variation of these eight parameters over the database. Therefore, we have not followed a strictly statistical approach and have instead developed scaling relations based partially on statistical arguments and partially on an analysis of how $\tau_E$ scales for parameter scans in different experiments and in different regimes. The primary reason for this is that some key parameters (such as B) were either almost the same for all of the data from the tokamaks in the database or did not vary much over the whole dataset (such as M).

With this approach, the ITER database has been analysed, and a class of scalings which are good fits to the data have been selected. The differences between these successful scalings [1] cannot be resolved within the present database (Table I, where $R^2$ denotes the fraction of the total (corrected) sum of squares that is explained by the model, see Ref. [8]). We found that this is due to two specific features of the ITER database: collinearity in the data due to a lack of variation in some of the directions in the multi-dimensional parameter space of the database $\tau_E = \tau_E(P,B,I_p,k,R,a,n,M)$, and variations in the scaling of $\tau_E$ from tokamak to tokamak [2-5]. We find that the most poorly defined parameter combination in terms of

the parameters that are varied least in the database is the shape index $f_S$:

$$f_S = 0.32 \, (R/a) \, a^{0.25} \, k^{0.5} \quad (1)$$

where R is the major radius in m, a is the minor radius in m, and k is the elongation at the null point (see, for example, Ref. [3]). In statistical terms, this is the least well determined principal component. The direction of least variation in the ITER database arises partly from the fact that tokamaks with circular cross-sections (e.g. ASDEX and JT-60) tend to have higher aspect ratios. The 'shape' factor $f_S$ tends to be about unity for almost all of the tokamaks in the database (Table II and Fig. 1), while the value of $f_S$ for ITER, for example, is about 1.6.

Another difficulty in the determination of scaling expressions is the variation in the scaling with respect to the machine parameters from tokamak to tokamak [2]. Part of the variation in the confinement scaling from tokamak to tokamak may be related to physics differences, which could be corrected if profile data such as the radial profile of the heating power deposition were available. The scaling with respect to the power is similar for most of the tokamaks in the database ($\tau_E \sim P^{-0.5 \pm 0.1}$). The scaling of $\tau_E$ with density n and plasma current $I_p$, however, varies more strongly from device to device [1, 3, 5, 6]. The dominant scaling of $\tau_E$ with $I_p$ is $I_p^{\alpha \pm \beta}$, with $\alpha \sim 1$ and $\beta \leq 0.2$. The variation in $\beta$ from tokamak to tokamak can be interpreted as being due to the variation in the scaling with respect to q. Most of the discharges in the dataset were limiter discharges and were heated with neutral beams. A number of radiofrequency heated and divertor discharges were included for comparison. To a large extent, the differences in confinement are negligible. However, we mention briefly two distinct groups of discharges which scaled differently. The incremental confinement time $\tau_{inc}$ depends strongly on $I_p$ in JT-60 limiter discharges, but depends only weakly on $I_p$ in JT-60 divertor discharges. A strong dependence of $\tau_E$ on density has been observed in T-10 electron cyclotron heating experiments, while a much weaker dependence of $\tau_E$ on density has been observed in tokamaks with neutral beam injection (NBI) heating. The scalings with current and density are linked because higher current operation is often accompanied by higher operating densities and therefore the density and current variations are correlated. These different dependences may also be due to different operating conditions and parameters that are not included in the database. We find that much of the variation in the scaling with current among tokamaks can be included by introducing a factor $f_q$ to reflect the variation with q (and implicitly $I_p$):

$$f_q = 1.56\, a^2\, kB/RI_p = q_{eng}/3.2 \qquad (2)$$

where B is the magnetic field in T, $I_p$ is the plasma current in MA, and $q_{eng} = 5\, a^2 kB/RI_p$ is the usual engineering cylindrical safety factor (3.2 is the average value of $q_{eng}$ for the ITER database). The lack of variation in $f_S$ in the existing database and the poorly determined dependence of $\tau_E$ on $q_{eng}$ are the main reasons for the corresponding uncertainties in the exponents in each scaling expression.

TABLE II. TYPICAL VALUES OF $f_S$

| Tokamak | R | a | A | k | $f_S$ |
|---|---|---|---|---|---|
| JT-60 | 3.0 | 0.9 | 3.3 | 1 | 1.0 |
| TFTR | 2.5 | 0.8 | 3.1 | 1 | 0.9 |
| JET | 2.9 | 1.2 | 2.4 | 1.45 | 0.9 |
| JFT-2M | 1.3 | 0.3 | 2.4 | 1.45 | 1.1 |
| DIII-D | 1.7 | 0.6 | 2.5 | 1.8 | 1.0 |
| DIII | 1.45 | 0.4 | 3.6 | 1.4 | 1.0 |
| ISX-B | 0.95 | 0.25 | 3.8 | 1.4 | 1.0 |
| DITE | 1.2 | 0.25 | 4.8 | 1 | 1.0 |
| ASDEX | 1.65 | 0.4 | 4.1 | 1 | 1.0 |
| TFR | 1.0 | 0.2 | 5 | 1 | 1.0 |
| PDX | 1.45 | 0.4 | 3.6 | 1 | 0.9 |
| PLT | 1.35 | 0.4 | 3.4 | 1 | 0.8 |
| T-10 | 1.5 | 0.3 | 5 | 1 | 1.1 |

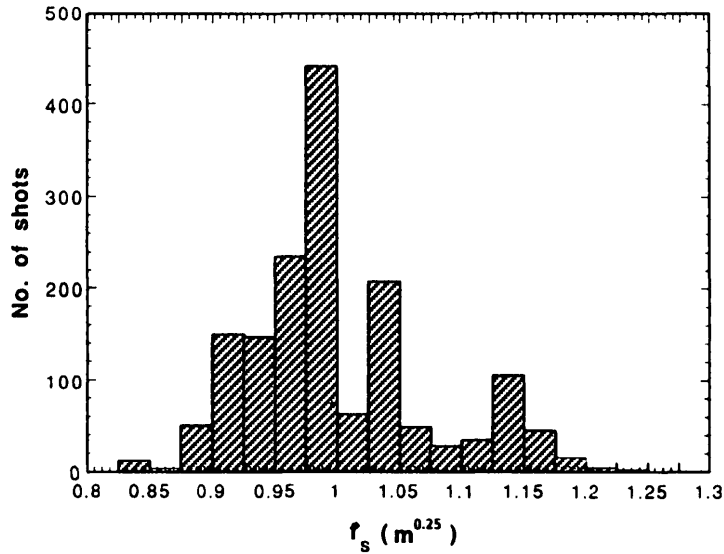

FIG. 1. *Number of shots for $f_S = 0.32\, R\, a^{-0.75}\, k^{0.5}$ in the ITER L-mode database [1].*

In the present ITER database, there is virtually no isotope variation within each tokamak, so that it cannot be determined by regression. We have assumed an isotope dependence proportional to $M^{0.5}$ on the basis of the results of the survey of isotope dependences conducted by Wagner et al. [12], which shows an enhancement factor of $\sim 1.4$ for the energy confinement time for operation with deuterium compared to hydrogen.

In Section 2, we propose a new power law scaling and show that many of the commonly accepted power law scaling expressions differ from each other by multiples of $f_S$ and $f_q$. In Section 3, the same methodology is applied to analyse offset-linear scaling expressions. Section 4 gives a summary and conclusions.

## 2. POWER LAW SCALING EXPRESSIONS

We have considered power law scaling expressions for $\tau_E$ constructed from eight basic tokamak operating parameters: M, $I_p$, R, a, k, n, B and P, of the form $\tau_E = C\, M^{\alpha_M}\, I_p^{\alpha_I}\, R^{\alpha_R}\, a^{\alpha_a}\, k^{\alpha_k}\, n^{\alpha_n} B^{\alpha_B}\, P^{\alpha_P}$, where $\tau_E$ is the global energy confinement time in seconds, M is the average isotopic mass number, $I_p$ is the plasma current in MA, R is the major radius in m, a is the minor radius in m, k is the elongation at the plasma surface, n is the line averaged plasma density in units of $10^{20}$ m$^{-3}$, B is the toroidal field in T and P is the total plasma heating power in MW; the $\alpha$'s are the exponents for the fit.

By combining a number of separate data analyses, we have determined the following scaling expression which fits all major groups of data:

$$\tau_E^{ITER89-P} = 0.048 \, M^{0.5} \, I_p^{0.85} \, R^{1.2} \, a^{0.3} \, k^{0.5} \, n^{0.1} \, B^{0.2} \, P^{-0.5} \quad (3)$$

The ITER scaling law differs from the previous scaling laws only slightly in the well determined parameter directions. In the poorly determined parameter directions, the ITER scaling was chosen to adequately fit all significant groups of discharges. The $I_p$ dependence in the ITER scaling was chosen as a compromise between the linear scaling observed in most tokamaks and the weaker scaling observed in JET and JT-60. As mentioned before, the vector $f_S$ is about 1 for the complete database, and the factor $f_q$ reflects the fact that the q dependence of $\tau_E$ is poorly determined because of tokamak to tokamak variations in the q scaling. We find that the power law scaling expressions that provide a reasonable fit to the experimental data can be expressed in the general form (see Refs [3, 4]):

$$\tau_E^{scaling} = \tau_E^{ITER89-P} \times f_S^{\alpha_S} \, f_q^{\alpha_q} \quad (4)$$

where the values of $\alpha_S$ and $\alpha_q$ are in the ranges $|\alpha_S| < 0.7$ and $|\alpha_q| < 0.2$. For example, by appropriate choices of $\alpha_S$ and $\alpha_q$, it is possible to reproduce the Kaye-Big scaling [1]:

$$\tau_E^{KB} = 0.105 \, M^{0.5} \, I_p^{0.85} \, R^{0.5} \, a^{0.8} \, k^{0.25} \, n^{0.1} \, B^{0.3} \, P^{-0.5}$$

$$= \tau_E^{ITER89-P} \times f_S^{-0.7} \, (0.98 \, a^{-0.02} \, k^{0.1} \, B^{0.1}) \quad (5)$$

the Goldston scaling [9]:

$$\tau_E^G = 0.03 \, M^{0.5} \, I_p \, R^{1.75} \, a^{-0.37} \, k^{0.5} \, P^{-0.5}$$

$$= \tau_E^{ITER89-P} \times f_S^{0.4} \, f_q^{-0.15} \, (1.05 \, a^{-0.07} \, k^{-0.05} \, n^{-0.1} \, B^{-0.05}) \quad (6)$$

and the scaling by Chudnovskij-Yushmanov [4]:

$$\tau_E^{CY} = 0.125 \, M^{0.5} \, I_p^{0.65} \, R^{0.4} \, a^{1.1} \, k^{0.3} \, n^{0.15} \, B^{0.35} \, P^{-0.5}$$

$$= \tau_E^{ITER89-P} \times f_S^{-0.6} \, f_q^{0.2} \, (1.2 \, a^{-0.05} \, k^{-0.1} \, n^{0.05} \, B^{-0.05}) \quad (7)$$

The uncertainty in $\alpha_S$ is larger than that in $\alpha_q$ because the variation of $f_S$ is much less than the variation of $f_q$ in the database. The most straightforward estimate for the uncertainties in $\alpha_S$ and $\alpha_q$ is given by ordinary least squares regression:

$$F = N^{-1} \sum_i [\ln(\tau_{E_i}/\tau_{E_i}^{fit})]^2$$

denotes the residual sum of squares for the scaling expression, where the sum over i is over all of the shots in the database. To leading order, F is a quadratic function of the regression coefficients. Neglecting the influence of the well determined directions, we can approximate F by

$$F(\alpha_S, \alpha_q) \approx 0.17^2 \, [1 + 0.15 \, (\alpha_S - \hat{\alpha}_S)^2 + 2.0 \, (\alpha_q - \hat{\alpha}_q)^2] \quad (8)$$

where $\hat{\alpha}_S$ and $\hat{\alpha}_q$ are the exact least squares regression values. For a given confidence level z, the standard $\chi^2$ test rejects all $\alpha_S$ and $\alpha_q$ values for which

$$F(\alpha_S, \alpha_q) \geq F(0,0) \, (1 + N^{-1} \, \chi^2_{2;z}) \quad (9)$$

where N is the total number of observations, and $\chi^2_{2;z}$ is the critical value, for a $\chi^2_2$ variate, corresponding to a tail probability z. From Eqs (8) and (9) we can estimate the uncertainty of the confinement time extrapolation for ITER. We use $d_S$ for the distance in the direction of $f_S$ from the database mean, i.e. $d_S = |\ln f_S^{ITER} - \langle \ln f_S \rangle|$, and similarly for $d_q$. The prediction half-width is then given by $[N^{-1} \, \chi^2_{2;z} \times (0.15^{-1} \, d_S^2 + 2^{-1} \, d_q^2)]^{1/2}$. For z = 5% ($\chi^2_{2;z} = 6$), N = 1200, $d_S = |\ln 1.6| = 0.47$ and $d_q = |\ln 0.59| = 0.525$, this half-width (two standard deviations) is about 10%.

Standard least squares regression analysis yields a half-width (two standard deviations) deviation for $\ln \tau_{E; ITER}$ of 14% (which means that the other regression variables contribute about 1/3 of the total interval). This interval is a sensible lower bound on the real uncertainty, since it neglects the tokamak to tokamak variation, and it assumes that a power law type scaling, which depends only on the variables included, is valid. The tokamak to tokamak variation has been investigated in Refs [2, 3]. An approximate formula [2, 4] can be obtained by replacing the number of data points N in Eq. (9) by the number of tokamaks. More accurate formulas are presented in Refs [2, 3], which give a half-width (two standard deviations) of about 40%.

For the ITER parameters: $M = 2.5$, $I_p = 22$ MA, $R = 6$ m, $a = 2.15$ m, $k = 2.21$, $n = 1.3 \times 10^{20}$ m$^{-3}$, $B = 4.9$ T, and $P = P_{OH} + P_{aux} + P_\alpha - P_{rad} = 150$ MW ($P_{rad}$ includes only bremsstrahlung and synchrotron radiation from the plasma centre), Eq. (3) yields the following L-mode energy confinement time prediction for ITER with a power law scaling:

$$\tau_E^{ITER89-P} = 1.95 \text{ s} \tag{10}$$

An approximate prediction interval of 66% (corresponding to one standard deviation) for the ITER L-mode confinement time $\tau_E^{ITER89-P}$ is 1.95 s × (1 ± 0.2).

## 3. OFFSET-LINEAR SCALINGS

We have also fitted the L-mode data with an offset-linear-type of scaling of the form $\tau_E = W_{OH}/P + \tau_{inc}$. We have compared three offset-linear scalings: those by Odajima-Shimomura [11], Rebut-Lallia [10] and Takizuka [5]. To develop a better Ohmic scaling, we used additional data in the saturated Ohmic regime from outside the ITER L-mode database, from JET, JT-60, TFTR, DIII-D, TEXTOR, DIII, ASDEX, T-10, JFT-2, PLT, JFT-2M and ALCATOR-C. The Ohmic stored energy $W_{OH}$ (MJ) for the Odajima-Shimomura scaling [11] is given by

$$W_{OH}^{OS} = 0.064 \, M^{0.5} \, I_p \, R^{1.6} \, a^{0.4} \, k^{0.2} \, n^{0.6} \, B^{0.2} \, f(Z_{eff}) \, g(q) \tag{11}$$

where $f(Z_{eff}) = Z_{eff}^{0.4} \, [(15 - Z_{eff})/20]^{0.6}$ and $g(q) = [3q(q + 5)/(q + 2)(q + 7)]^{0.6}$. For the Rebut-Lallia scaling [10] we have

$$W_{OH}^{RL} = 0.17 \, M^{0.5} \, I_p^{0.5} \, L^{2.75} \, n^{0.75} \, B^{0.5} \, Z_{eff}^{0.25} \tag{12}$$

where $L^3 = R \, a^2 k$ and the ratio of total stored energy to electron energy is assumed to be 1.65. Another expression for $W_{OH}$ has been proposed recently by Takizuka [5]:

$$W_{OH}^T \sim M^{0.2} \, I_p^{0.8} \, R^{1.6} \, a^{0.6} \, k^{0.5} \, n^{0.6} \, B^{0.35} \, Z_{eff}^{0.4} \tag{13}$$

Except for the mass and $Z_{eff}$ dependence and the weak factor $g(q)$, these expressions can be made very similar by multiplying them by various powers of $f_S$ and $f_q$. The Odajima-Shimomura scaling can be obtained from the Takizuka scaling by $W_{OH}^{OS} \sim W_{OH}^T \times f_S^{-0.2} \, f_q^{-0.2}$.

To make the Rebut-Lallia and the Takizuka scalings agree, a small power of $na^2/I_p$ is necessary in addition to factors of various powers of $f_S$ and $f_q$: $W_{OH}^{RL} \sim W_{OH}^T \times (na^2/I_p)^{0.15} \, f_S^{-0.5} \, f_q^{0.2}$. Because of the small deviations of $f_S$, $f_q$ and $na^2/I_p$ in the ITER database, all these expressions can reproduce the experimental data reasonably well when $\tau_{inc} \ll W_{OH}/P$ [4, 5]. Among the three forms of $W_{OH}$, Eq. (12) is preferred because it has the smallest deviation of $\tau_{inc}$ from the power law form [4, 5]. Neglecting the $Z_{eff}$ dependence of $W_{OH}^T$ and determining the constant to fit the database, the basic expression for $W_{OH}$ is given by

$$W_{OH}^{ITER89} = 0.064 \, M^{0.2} \, I_p^{0.8} \, R^{1.6} \, a^{0.6} \, k^{0.5} \, n^{0.6} \, B^{0.35} \tag{14}$$

The incremental confinement time $\tau_{inc}$ for the Odajima-Shimomura scaling

$$\tau_{inc}^{OS} = 0.085 \, M^{0.5} \, a^2 k \tag{15}$$

and for the Rebut-Lallia scaling

$$\tau_{inc}^{RL} = 0.014 \, M^{0.5} \, I_p \, R^{0.5} \, a \, k^{0.5} \, Z_{eff}^{-0.5} \tag{16}$$

which are based on fits to the JT-60/JFT-2M and JET data, respectively, give relatively poor fits to the whole experimental database [5, 6]. These two fits are only accurate when the Ohmic term $W_{OH}$ is dominant. On the basis of our analyses, we recommend the following form for $\tau_{inc}$:

$$\tau_{inc}^{ITER89} = 0.04 \, M^{0.5} \, I_p^{0.5} \, R^{0.3} \, a^{0.8} \, k^{0.6} \tag{17}$$

This new form of $\tau_{inc}$ provides a better fit to the whole experimental database than do the former scalings.

Analysis of the ITER database shows that, in addition, $a^2/I_p$ at fixed q is almost constant, and the size dependences of $\tau_{inc}^{OS}$ and $\tau_{inc}^{ITER89}$ are almost identical. However, for future devices, $I_p$ will be more nearly proportional to the size, and the dependence of $\tau_{inc}$ on the 'effective size' L will be different for each scaling: $\tau_{inc}^{OS} \sim L^2$, $\tau_{inc}^{RL} \sim L^{2.5}$, $\tau_{inc}^{ITER89} \sim L^{1.6}$.

The predicted value of $\tau_E$ using Eqs (14) and (17) is

$$\tau_E^{ITER89-OL} = W_{OH}/P + \tau_{inc} = 0.5 \text{ s} + 1.5 \text{ s} = 2.0 \text{ s} \tag{18}$$

which is nearly equal to the value of Eq. (10) given by $\tau_E^{ITER89-P}$.

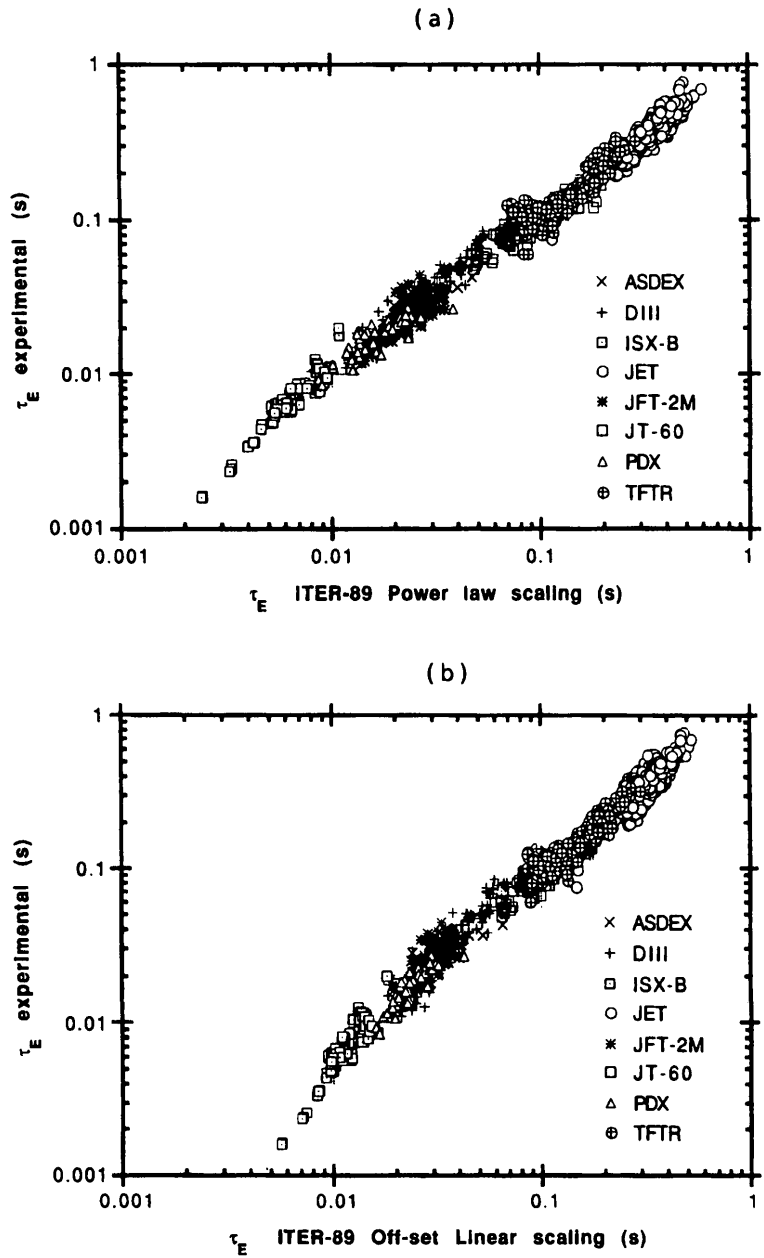

FIG. 2. Comparison of the scalings $\tau_E^{ITER89-P}$ (Eq. (19)) (with $\tau_E^{ITER89-OL}$ (Eq. (20))).
(a) Experimental $\tau_E$ versus $\tau_E$ from the ITER-89 power law scaling;
(b) Experimental $\tau_E$ versus $\tau_E$ from the ITER-89 offset-linear scaling.

## 4. CONCLUSIONS AND DISCUSSION

The present multiplicity of scaling expressions for $\tau_E$ is due both to the lack of variation in $f_S \sim R^{-0.75} k^{0.5}$ for the present tokamak devices and to variation in the dependence of $\tau_E$ on the physical parameters among different tokamaks. On the basis of the ITER database of L-mode energy confinement, we have developed two new scalings as a basis for the ITER design. The recommended power law scaling is:

$$\tau_E^{ITER89-P} = 0.048 \, M^{0.5} \, I_p^{0.85} \, R^{1.2} \, a^{0.3} \, k^{0.5} \, n^{0.1} \, B^{0.2} \, P^{-0.5} \tag{19}$$

and the offset-linear scaling is:

$$\tau_E^{ITER89-OL} = 0.064 \, M^{0.2} \, I_p^{0.8} \, R^{1.6} \, a^{0.6} \, k^{0.5} \, n^{0.6} \, B^{0.35}/P$$
$$+ 0.04 \, M^{0.5} \, I_p^{0.5} \, R^{0.3} \, a^{0.8} \, k^{0.6} \tag{20}$$

A comparison of the predicted confinement scalings with the experimental data is presented in Fig. 2.

The differences between the power law scaling and the offset-linear scaling are difficult to distinguish for moderate values of $P/P_{OH}$. We note that the offset-linear scaling requires more free parameters and is therefore less well determined in the present ITER database. To determine a more accurate offset-linear scaling, the present database must be supplemented by more data with a broader range of $P_{aux}/P_{OH}$ values. Thus, while the present ITER89-OL scaling fits the data nearly as well as the ITER89-P scaling, and while both types of scaling give approximately the same confinement time prediction for ITER parameters:

$$\tau_E^{ITER89-P} = 2.0 \text{ s} \times (1 \pm 0.2) \qquad (21)$$

the predictive ability of the offset-linear scaling is judged to be less. Therefore, we recommend that the ITER89-P scaling be used for design extrapolations and that the ITER89-OL scaling be used only as a reference point.

The dependences of $\tau_E$ on n, $I_p$, M and B have to be investigated more thoroughly, and the physical reasons for the tokamak to tokamak variation of some of the dependences have to be clarified. The differences in the density scaling for different tokamaks may be due to the dependence of the radial profile of the power deposition for different heating methods. The weak dependence of $\tau_E$ on the plasma density n in beam heated tokamaks, which comprise most of the points in the database, may be attributed to a decreased beam penetration at higher densities and to an increased superthermal ion contribution at low density. Therefore, the dependence on density in the scaling obtained from the database analysis may have a systematic error which would lead to an underestimate of the confinement time that would be obtainable with strong central heating such as alpha particle heating in high Q and ignited tokamaks. In addition, there is a systematic correlation of $I_p$ with n in many tokamak experiments.

To increase the accuracy of the prediction of $\tau_E$, new data with increased variation in the shape index, $f_S = 0.32 \text{ R a}^{-0.75} \text{ k}^{0.5}$, are necessary. Such experiments have been performed in JT-60 and TFTR and support the ITER89-P scaling surprisingly well. When additional data with more $f_S$ values become available, they will be analysed and the appropriate modifications to the ITER89 scaling will be made in order to reduce the uncertainties and to increase the accuracy of the predictions.

The equations which describe local transport in a turbulent plasma place a constraint on the power law scaling [13] which is of the form

$$5\alpha_B + \alpha_I + 8\alpha_n + 3\alpha_P - 4\alpha_R - 4\alpha_a + 5 = 0 \qquad (22)$$

In situations where the heating profile is a function of density, as is the case with NBI, as mentioned above, or where the atomic physics in the edge region plays a dominant role in the confinement, these simple constraints will no longer hold. Interestingly, we find that Eq. (19) is close to satisfying the constraint imposed by Eq. (22). Only a change of the n exponent from $\alpha_n = 0.1$ to $\alpha_n = 0.08$ is necessary, and this is within the error bars. Scaling (19) satisfies even better the non-linear gyrokinetic equations [14] (NGK). Minor changes of the indices of Eq. (19) to

$$\tau_E \propto I_p^{0.9} R^{1.3} a^{0.4} k^{0.5} B^{0.2} P^{-0.4} \qquad (23)$$

yield a form which satisfies the constraints of the NGK equation. The resulting thermal diffusivity $\chi$ has the form

$$\chi \propto (\rho^2 v_{th}/L) \nu_*^{0.4} \qquad (24)$$

where $\rho$ is the Larmor radius, $v_{th}$ is the thermal velocity, L is a characteristic scale length and $\nu_*$ is the ratio of the collision frequency to the bounce frequency.

We have not derived a systematic scaling for H-mode confinement because of the lack of a complete database, and we recognize that confinement with H-mode operation will probably have dependences that are not present in the L-mode experiments. Nonetheless, a comparison of the ITER89-P scaling, Eq. (19), with the H-mode data from ASDEX and JET indicates that the H-mode confinement time in ASDEX and JET is roughly 2.3 and 2.1 times, respectively, the confinement time predicted by the ITER89 power law scaling. Currently, there are activities to collect a systematic database for the H-mode [15], and when these data are available, a scaling for the H-mode will be developed.

## ACKNOWLEDGEMENTS

The database used in the paper has been collected from JT-60, TFTR, JET, T-10, JFT-2M, DIII-D, DIII, ISX-B, DITE, ASDEX, TFR, PDX and PLT. The authors are grateful to the members of the experimental teams of these tokamaks for their contributions to the ITER database and for the careful selection of reliable data in a form that made these analyses possible.

The authors would also like to express their sincere thanks to all of the participants of the ITER Confinement Workshop, 31 July to 11 August 1989, at the IPP, Garching, for fruitful discussions. They are grateful especially to Drs. V. Mukhovatov and F. Engelmann for their continuous encouragement and support.